# Neutron Scattering Studies of the Breathing Pyrochlore Antiferromagnet LiGaCr$_4$O$_8$

Zheng He,[1] Yiqing Gu,[1,2] Hongliang Wo,[1,2] Yu Feng,[1,3,4] Die Hu,[1] Yiqing Hao,[1] Yimeng Gu,[1,2] Helen C. Walker,[5] Devashibhai T. Adroja,[5,6] and Jun Zhao[1,7,2,8,9,*]

[1]*State Key Laboratory of Surface Physics and Department of Physics, Fudan University, Shanghai 200433, China*

[2]*Shanghai Qi Zhi Institute, Shanghai 200232, China*

[3]*Institute of High Energy Physics, Chinese Academy of Sciences (CAS), Beijing 100049, China*

[4]*Spallation Neutron Source Science Center (SNSSC), Dongguan 523803, China*

[5]*ISIS Facility, Rutherford Appleton Laboratory, STFC, Chilton, Didcot, Oxon OX11 0QX, United Kingdom*

[6]*Highly Correlated Matter Research Group, Physics Department, University of Johannesburg, P.O. Box 524, Auckland Park 2006, South Africa*

[7]*Institute of Nanoelectronics and Quantum Computing, Fudan University, Shanghai 200433, China*

[8]*Shanghai Research Center for Quantum Sciences, Shanghai 201315, China*

[9]*Collaborative Innovation Center of Advanced Microstructures, Nanjing, 210093, China*

Abstract:

We report neutron scattering measurements of the spinel oxide LiGaCr$_4$O$_8$, in which magnetic ions Cr$^{3+}$ form a breathing pyrochlore lattice. Our experiments reveal the coexistence of a nearly dispersionless resonance mode and dispersive spin wave excitations in the magnetically ordered state, which can be quantitatively described by a quantum spin model of hexagonal loops and linear spin wave theory with the same set of exchange parameters, respectively. Comparison to other Cr spinel oxides reveals a linear relationship between the resonance energy and lattice constant across all these materials, which is in agreement with our hexagonal loop calculations. Our results suggest a unified picture for spin resonances in Cr spinel oxides.

Geometrical frustration occurs in lattices based on triangles or tetrahedra, where magnetic exchange interactions between local moments cannot be simultaneously satisfied. Geometrical frustration may prevent long-range magnetic order and lead to highly degenerated ground states with exotic spin-spin correlations and emergent excitations [1]. Different correlated behaviors in frustrated magnets give rise to various magnetic ground states, which are of particular interest across many fields of physics [2,3].

The spinel oxide ACr$_2$O$_4$ (A = Zn, Mg, Hg...) is a canonical example of geometrically frustrated magnets [4,5]. Transition metal Cr atoms, which are octahedrally coordinated by oxygen atoms, form a pyrochlore lattice with corner-sharing tetrahedra. The pyrochlore lattice consists of alternating stacked kagome and triangle planes along the [111] direction and is highly frustrated. The Cr spinel oxides in general order at temperatures ($T_N$) much lower than the Curie-Weiss temperatures owing to strong frustration [5]. Interestingly, the most prominent feature in the spin excitation spectrum of Cr spinel oxides is a nearly dispersionless resonance mode in the magnetically ordered state [6-10]. For instance, strong resonance excitations were found in the ordered state of $MgCr_2O_4$ and phenomenologically explained by hexamer excitations [7,8], similar to the diffusive scattering observed in the paramagnetic state. However, a recent work instead suggested that the spin resonances in $MgCr_2O_4$ are associated with heptamers [11]. As yet, a unified description of the resonance mode in spinel oxides remains lacking. There is also no consensus on the underlying magnetic exchange interaction that drives the resonance mode among different classes of Cr spinel oxides.

The A-site ordered Cr spinel oxide $LiGaCr_4O_8$ offers a new opportunity to study the nature of spin excitations and magnetic interactions in spinel oxides [12,13]. In $LiGaCr_4O_8$, the magnetic ions $Cr^{3+}$ [$(t_{2g})^3$ with spin $S = 3/2$] form alternating arrays of small and large tetrahedra (breathing pyrochlore lattice). $LiGaCr_4O_8$ exhibits an antiferromagnetic phase transition at $T_N \approx 14$ K, which is much lower than the Curie-Weiss temperature $\theta_{cw}$ of -659 K [12]. This suggests that even with alternation of small and large tetrahedra, the system shows strong frustration. Similar to other uniform Cr spinel oxides, recent neutron diffraction measurements on $LiGaCr_4O_8$ revealed multiple magnetic propagation vectors [$k = (0, 0, 1)_c$ and $k = (1/2, 1/2, 1/2)_t$] below $T_N$, accompanied by the small portion of samples transforming into a low-symmetry tetragonal phase [14,15]. The total ordered moment of $LiGaCr_4O_8$ is $M \sim$ 1.39 $\mu_B/Cr^{3+}$ averaged over the whole sample, which is significantly lower than the fully ordered moment of 3 $\mu_B$ for the $Cr^{3+}$ spin [15]. This indicates magnetic frustration is only partially relieved and strong spin fluctuations persist even in the ordered phase.

To investigate the spin excitations in $LiGaCr_4O_8$, we performed inelastic neutron scattering experiments on the MERLIN direct geometry chopper spectrometer at the ISIS spallation neutron source (U.K.) [16]. We chose incident neutron energies of 16.1, 35.3 and 50.6 meV with energy resolutions of 0.89, 1.29 and 4.23 meV at the elastic line, respectively. High-quality polycrystalline $LiGaCr_4O_8$ was synthesized by the solid-state reaction method as introduced in Ref. [12]. Approximately 39.7 g of powder sample was loaded into an annular aluminum foil packet inside of an aluminum can with a diameter of 40 mm to minimize the strong neutron absorption of Li nuclei. The whole assembly was mounted on the cold head of a closed-cycle $^4$He refrigerator to reach the base temperature of 5 K. Collected inelastic neutron data were analyzed using the MantidPlot software [17]. Spin wave simulations were carried out using SpinW software packages in the Holstein-Primakoff approximation and the powder average was performed with 20,000 random orientations distributed over the Brillouin zone at each scattering vector $Q$ [18].

Figure 1(a) shows the inelastic neutron scattering spectrum in the paramagnetic phase. Diffusive magnetic excitations with the maximum intensity at around $Q = 1.5$ Å$^{-1}$ are widely distributed in energy ($E$) and momentum ($Q$) space. The $Q$ dependence of quasielastic excitations in a constant energy cut shows a rather broad peak, providing evidence for short-range antiferromagnetic (AFM) spin correlations [Fig. 1(b)]. This result resembles those observed in other spinel oxides [6,7,10,19-24].

Now, we turn to spin excitations in the magnetically ordered state. Figures 2(a)

and 2(b) show inelastic neutron scattering data at 5 K with different incident neutron energies. It is clearly revealed that the majority of the spectral weight in the quasielastic region [Fig. 1(a)] shifts to higher energies, forming a strong dispersionless resonance mode at ~ 6 meV and weaker dispersive spin-wave like excitations near $Q \approx 1.5 \sim 1.7$ Å$^{-1}$. In order to understand the resonance mode at finite energy in the magnetically ordered state quantitatively, we adopt a quantum hexagonal spin loop model, which extends from the idea of the classical hexamer model [7,8,25] appropriate for quasielastic scattering in the paramagnetic state [3]. As shown in Fig. 2(f), the hexagonal loop in LiGaCr$_4$O$_8$ is formed by alternating bonds with bond lengths $r$ and $r'$ ($r < r'$), which correspond to nearest-neighbor (NN) AFM interactions $J$ and $J'$ ($J > J' > 0$) respectively. Since NN couplings are predominant and there is no evidence for the spin anisotropy [26,27], the Hamiltonian of this Heisenberg spin loop can be given by

$$\widehat{H} = J \sum_{i=1,3,5} (\mathbf{S}_i \cdot \mathbf{S}_{i+1} + B_f \mathbf{S}_{i+1} \cdot \mathbf{S}_{i+2}) \qquad (1)$$

where $B_f = J'/J$ is defined as the breathing factor, $\mathbf{S}_i$ denotes the spin operator of the $i$-th spin on the loop and $\mathbf{S}_7 \equiv \mathbf{S}_1$ is the periodic boundary condition for the hexagonal structure. The Hamiltonian matrix is built up by the product states with spin number $n$ = 6 and spin length $S$ = 3/2, thus giving rise to $(2S+1)^n$ = 4096 basis states. Although the large dimension of the Hilbert space makes numerical simulations difficult, we have designed a parallel algorithm in a high-performance computing cluster to reduce the computational complexity and make quantitative analyses easier in the following context. First of all, the accurate values of $J$ and $J'$ are required for setting up the Hamiltonian of this bipartite spin loop. These two parameters can be extracted from the resonance energy with a useful constraint as introduced below. By the exact diagonalization of the Hamiltonian (1), the calculated resonance energies as the function of the AFM interaction $J$ and the breathing factor $B_f$ are plotted in Fig. 2(d). The black curve is the 6.0 meV contour line of the calculated energy map, which is based on the resonance energy from our neutron scattering measurements. The white curve is the constraint between $J$ and $J'$ with $dJ = J(1 - B_f) \approx 4.1$ meV, which is obtained by multiplying (i) the empirical relation of $dJ/dr \approx -40$ meV/Å in chromium oxides [6,9,28] and (ii) the difference between bond lengths $r$ and $r'$ in cubic phase LiGaCr$_4$O$_8$ with $dr = r - r' = -0.103$ Å [12]. The intersection of these two curves yields the magnetic interactions $J = 10.4$ meV and $J' = 6.2$ meV with the breathing factor $B_f = 0.60$. We note that the $B_f$ derived from our experiment is quite close to the values estimated from classical Monte Carlo simulations ($B_f = 0.6$) and recent GGA+$U$ calculations ($B_f = 0.66$) [12,27]. Furthermore, the values of $J$ and $J'$ are just slightly higher than the calculated result ($J = 8.6$ meV and $J' = 5.7$ meV) from density functional theory. The small deviation is probably because the first principle calculation is based on a larger lattice parameter at room temperature [27].

With the $J$ and $B_f$ given above, the spin Hamiltonian in Eq. (1) can be numerically resolved and the low energy section of the calculated spectrum is displayed in Fig. 2(e). From the perspective of spin loop excitations, these discrete energy levels are classified into $L$- and $E$- band according to their shift quantum numbers [29] and the observed spin resonance in LiGaCr$_4$O$_8$ is the excitation from the singlet ground state $|\psi_L(S=0)\rangle$ to the triplet first-excited state $|\psi_L(S=1)\rangle$ labeled as $L_0$ in the figure. Since eigenstates $|\psi_L(S=0)\rangle$ and $|\psi_L(S=1)\rangle$ have been obtained through the exact diagonalization, the momentum distribution of the resonance mode can be calculated by the dynamical structure factor as follows [30]

$$S(\mathbf{Q}, \omega) = C|F(Q)|^2 \sum_{ij} e^{i\mathbf{Q} \cdot \mathbf{R}_{ij}} \sum_{\alpha\beta} \left(\delta_{\alpha\beta} - \frac{Q_\alpha Q_\beta}{Q^2}\right)$$
$$\times \sum_{\psi\psi'} \langle\psi|\hat{s}_{i\alpha}|\psi'\rangle\langle\psi'|\hat{s}_{j\beta}|\psi\rangle \delta(\hbar\omega + E_\psi - E_{\psi'}) \quad (2)$$

where $C$ is the scale factor, $F(Q)$ is the magnetic form factor of $Cr^{3+}$ [31], $\mathbf{R}_{ij} = \mathbf{R}_i - \mathbf{R}_j$ with the spin position vector $\mathbf{R}_i$, $(\delta_{\alpha\beta} - \frac{Q_\alpha Q_\beta}{Q^2})$ denotes the orientation factor with $\alpha$, $\beta = x, y$ and $z$ and $\hat{s}_{i\alpha}$ is the spin operator of the $i$-th spin in the direction $\alpha$. After averaging four orientations of hexagonal loops on the lattice and all directions of the momentum transfer in reciprocal space, the calculated $Q$ dependence is in excellent agreement with the constant $E$ cuts of the spin resonance [Fig. 2(c)]. Therefore, we quantitatively show that the resonance mode in $LiGaCr_4O_8$ originates from hexagonal spin loops consisting of alternating bonds, with $J$ and $J'$ equal to 10.4 meV and 6.2 meV, respectively. Moreover, through the exact diagonalization method, we also attempted to fit the spin resonance mode using the quantum spin model of $S = 3/2$ heptamers (two corner-sharing tetrahedra). However, both the resonance energy and the integrated-momentum dependence obtained from this model are inconsistent with our neutron data (see Supplemental Material Sec. S1).

In addition to the resonance mode, $LiGaCr_4O_8$ exhibits weaker but clear spin wave like excitations with a columnar dispersion [Fig. 2(a)]. The dispersion of the spin wave excitations can be seen more clearly with a higher incident neutron energy of 35.3 meV [Fig. 3(a)]. It is shown that the columnar dispersion broadens with increasing energy and extends up to about 27 meV, which is consistent with constant $Q$ scans in Fig. 2(b).

To describe the wave vector and energy dependence of the spin-wave excitations, we adopt the Heisenberg $J$-$J'$ model with the same nearest exchange couplings determined in our spin loop model. Two magnetic orders identified in Ref. [15] are included in our simulation (see Supplemental Material Sec. S2). Figure 3(c) presents the calculated spin-wave spectrum, which well captures the feature of columnar dispersions emanating from magnetic Bragg points around $Q = 1.5$ Å$^{-1}$ but leaves the spin resonance at 6.0 meV with larger spectral weight unsolved. We further consider hexagonal spin loops' excitations [Fig. 3(d)] calculated by the dynamical structure factor in Eq. (2) and combine their contributions with simulated spin wave spectra. The total simulated scattering spectra displayed in Fig. 3(b) are in good agreement with our experimental data [Fig. 3(a)]. The fact that both types of spin excitations can be accurately described by the same sets of exchange parameters further verifies the validity of our quantum spin loop model.

In order to further elucidate the interplay between the spin resonance mode and spin wave excitations, we measured their detailed temperature evolution. Interestingly, the resonance mode and spin waves exhibit almost identical order-parameter-like temperature dependence and show a kink at $T_N$ [Fig. 3(e)]. The simultaneous development of spin-liquid-like loop excitations and spin wave excitations below $T_N$ suggests that the stable magnetic order not only promotes the long-range spin wave propagation mode, but also meets the antiparallel conditions inside the hexagonal loop, which leads to spin loop excitations. Above $T_N$, since the spin collinearity on the loop is destroyed by thermal fluctuations, the resonance excitation within a single hexagon diminishes and is replaced by broad diffusive excitations [Fig. 1(a)].

It is interesting to compare the resonance mode of $LiGaCr_4O_8$ with those of other uniform Cr spinel oxides. Based on our hexagonal loop model, the resonance energy $E_R$ of $ACr_2O_4$ can be directly extracted from the calculated map in Fig. 2(d) with the $B_f$ = 1, yielding a relation of $E_R = 0.705J$. By considering the empirical ratio of $dJ/dr \approx$ -40 meV/Å [6,9,28], our calculation suggests a linear relationship between the resonance energy $E_R$ and the lattice parameter $a$ with $dE_R/da$ = -10 meV/Å in uniform spinel oxides ($a = 2\sqrt{2}r$ in spinel lattices). As is shown in Fig. 4, the observed resonance energies and lattice parameters of $ACr_2O_4$ (A = Zn, Mg, Cd and Hg) indeed precisely follow this linear relationship [6-10,32,33]. This strongly indicates that the resonance modes in all other $ACr_2O_4$ are also associated with the hexagonal spin loops and the resonance energy provides an accurate measurement of the magnetic exchange interaction. Moreover, our calculation suggests that the resonance energy in $LiGaCr_4O_8$ would slightly deviate from the linear relationship owing to the presence of two different $J$ and $J'$ (with $B_f$ = 0.60). This also agrees with the data (Fig. 4). As for another A-site ordered spinel oxide $LiInCr_4O_8$, the breathing factor is as large as $B_f \sim 0.1$, meaning that the small tetrahedron is much more isolated than those of the uniform spinel oxides [12]. As a result, the excitation within a single tetrahedron would be favored rather than the hexagonal spin loop [34]. The observation of resonance mode in $LiInCr_4O_8$ at a higher energy than hexagonal loop calculation is consistent with this analysis (Fig. 4).

In summary, we have performed inelastic neutron scattering measurements of spin excitations in the A-site ordered spinel oxide $LiGaCr_4O_8$. Our data reveal the coexistence of a dispersionless resonance mode and dispersive spin wave excitations in the magnetically ordered state. We find that the strong spin resonance is associated with the hexagonal loops' excitations based on the quantum spin loop model while weak spin-wave excitations can be well reproduced by LSWT calculations. Both types of spin excitations can be described quantitatively with the same set of exchange coupling constants ($J$ = 10.4 meV, $J'$ = 6.2 meV), which is close to the values reported in a recent *ab initio* calculation [27]. In addition, we show that the resonance energy and lattice constant (Cr-Cr bond distance) follow a linear relationship among different classes of Cr spinel oxides, indicating that the resonance modes in these compounds have the same microscopic origin. The establishment of the correct quantum spin model and the determination of the dominant magnetic interactions provide the basis from which other exotic properties in similar systems could be understood.

Acknowledgements:

This work was partially supported by the National Key R&D Program of the MOST of China (Grant No. 2016YFA0300203), the Innovation Program of Shanghai Municipal Education Commission (Grant No. 2017-01-07-00-07-E00018), the Shanghai Municipal Science and Technology Major Project (Grant No. 2019SHZDZX01), and the National Natural Science Foundation of China (Grant No.11874119). The datasets for the neutron scattering experiment on the time-of-fight MERLIN spectrometer are available from the ISIS facility, Rutherford Appleton Laboratory data portal (10.5286/ISIS.E.RB1620075). All other data that support the plots within this Letter and other findings of this study are available from the corresponding author upon reasonable request.

* To whom all correspondence should be addressed.

zhaoj@fudan.edu.cn

References:

[1] L. Balents, Nature (London) **464**, 199 (2010).

[2] H. Diep *et al., Frustrated Spin Systems* (World Scientific, Singapore, 2013).

[3] C. Lacroix, P. Mendels, and F. Mila, *Introduction to Frustrated Magnetism: Materials, Experiments, Theory,* Vol. 164 (Springer, Berlin, 2011).

[4] S. -H. Lee *et al.,* J. Phys. Soc. Jpn. **79**, 011004 (2010).

[5] H. Ueda, H. Mitamura, T. Goto, and Y. Ueda, Prog. Theor. Phys. Suppl. **159,** 256 (2005).

[6] S. -H. Lee, C. Broholm, T. H. Kim, W. Ratcliff, II, and S. W. Cheong, Phys. Rev. Lett. **84**, 3718 (2000).

[7] K. Tomiyasu, H. Suzuki, M. Toki, S. Itoh, M. Matsuura, N. Aso, and K. Yamada, Phys. Rev. Lett. **101**, 177401 (2008).

[8] K. Tomiyasu, T. Yokobori, Y. Kousaka, R. I. Bewley, T. Guidi, T. Watanabe, J. Akimitsu, and K. Yamada, Phys. Rev. Lett. **110**, 077205 (2013).

[9] J. H. Chung, M. Matsuda, S. -H. Lee, K. Kakurai, H. Ueda, T. J. Sato, H. Takagi, K. P. Hong, and S. Park, Phys. Rev. Lett. **95**, 247204 (2005).

[10] K. Tomiyasu, H. Ueda, M. Matsuda, M. Yokoyama, K. Iwasa, and K. Yamada, Phys. Rev. B **84**, 035115 (2011).

[11] S. Gao *et al.,* Phys. Rev. B **97**, 134430 (2018).

[12] Y. Okamoto, G. J. Nilsen, J. P. Attfield, and Z. Hiroi, Phys. Rev. Lett. **110**, 097203

(2013).

[13] J.-C. Joubert and A. Durif, Bull. Soc. Fr. Mineral. **89**, 26 (1966).

[14] R. Wawrzynczak, Y. Tanaka, M. Yoshida, Y. Okamoto, P. Manuel, N. Casati, Z. Hiroi, M. Takigawa, and G. J. Nilsen, Phys. Rev. Lett. **119**, 087201 (2017).

[15] R. Saha, F. Fauth, M. Avdeev, P. Kayser, B. J. Kennedy, and A. Sundaresan, Phys. Rev. B **94**, 064420 (2016).

[16] R. I. Bewley, R. S. Eccleston, K. A. McEwen, S. M. Hayden, M. T. Dove, S. M. Bennington, J. R. Treadgold, and R. L. S. Coleman, Physica B (Amsterdam) **385B**, 1029 (2006).

[17] O. Arnold *et al.,* Nucl. Instrum. Methods Phys. Res., Sect. A **764**, 156 (2014).

[18] S. Toth and B. Lake, J. Phys. Condens. Matter **27**, 166002 (2015).

[19] K. Tomiyasu, H. Hiraka, K. Ohoyama, and K. Yamada, J. Phys. Soc. Jpn. **77**, 124703 (2008).

[20] K. Tomiyasu *et al.,* Phys. Rev. B **84**, 054405 (2011).

[21] A. Krimmel, H. Mutka, M. M. Koza, V. Tsurkan, and A. Loidl, Phys. Rev. B **79**, 134406 (2009).

[22] Y. Tanaka, R. Wawrzyńczak, M. D. Le, T. Guidi, Y. Okamoto, T. Yajima, Z. Hiroi, M. Takigawa, and G. J. Nilsen, J. Phys. Soc. Jpn. **87,** 073710 (2018).

[23] X. Bai *et al.,* Phys. Rev. Lett. **122**, 097201 (2019).

[24] G. Pokharel *et al.,* Phys. Rev. Lett. **125**, 167201 (2020).

[25] S. H. Lee, C. Broholm, W. Ratcliff, G. Gasparovic, Q. Huang, T. H. Kim, and S. W. Cheong, Nature (London) **418**, 856 (2002).

[26] S. Lee *et al.*, Phys. Rev. B **93**, 174402 (2016).

[27] P. Ghosh, Y. Iqbal, T. Muller, R. T. Ponnaganti, R. Thomale, R. Narayanan, J. Reuther, M. J. P. Gingras, and H. O. Jeschke, npj Quantum Mater. **4**, 63 (2019).

[28] K. Motida and S. Miyahara, J. Phys. Soc. Jpn. **28**, 1188 (1970).

[29] O. Waldmann, Phys. Rev. B **65**, 024424 (2001).

[30] A. Furrer and O. Waldmann, Rev. Mod. Phys. **85**, 367 (2013).

[31] P. J. Brown, in *International Tables for Crystallography* (Kluwer Academic Publishers, Dordrecht, 2004), Vol. C, Chap. Magnetic Form Factors, pp. 454–460.

[32] L. Ortega-San-Martin, A. J. Williams, C. D. Gordon, S. Klemme, and J. P. Attfield, J. Phys. Condens. Matter **20**, 104238 (2008).

[33] H. Ueda, H. Mitamura, T. Goto, and Y. Ueda, Phys. Rev. B **73**, 094415 (2006).

[34] G. J. Nilsen, Y. Okamoto, T. Masuda, J. Rodriguez-Carvajal, H. Mutka, T. Hansen, and Z. Hiroi, Phys. Rev. B **91**, 174435 (2015).

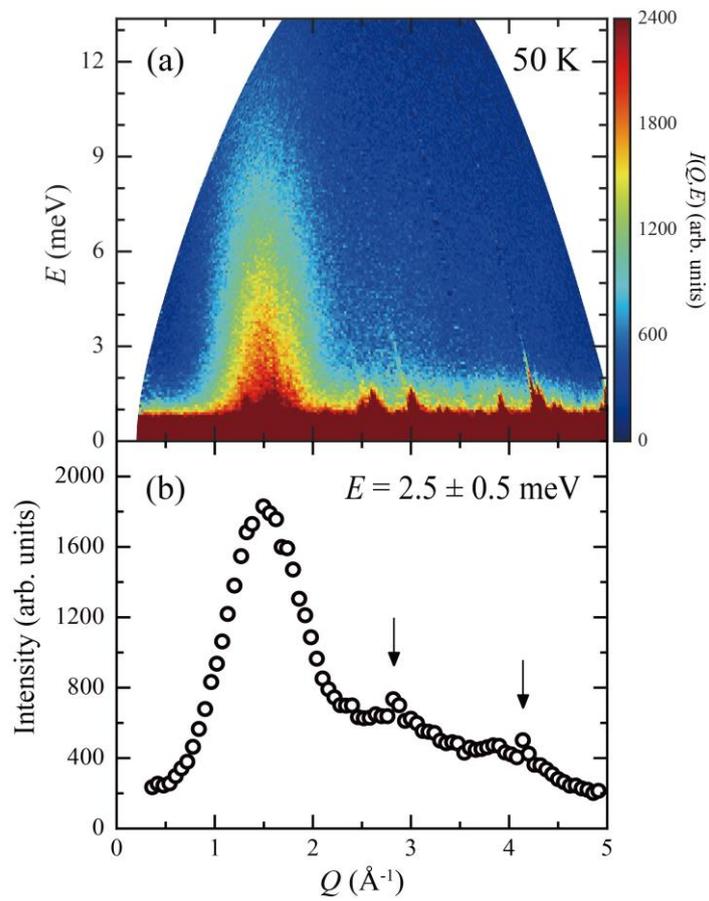

Fig. 1 (Color Online) (a) Magnetic excitation spectra of LiGaCr$_4$O$_8$ at 50 K with incident neutron energy $E_i$ = 16.1 meV. (b) The $Q$ dependence of magnetic excitations with the integrated energy from 2 to 3 meV. The arrows indicate the strong nuclear Bragg peaks (400) and (440).

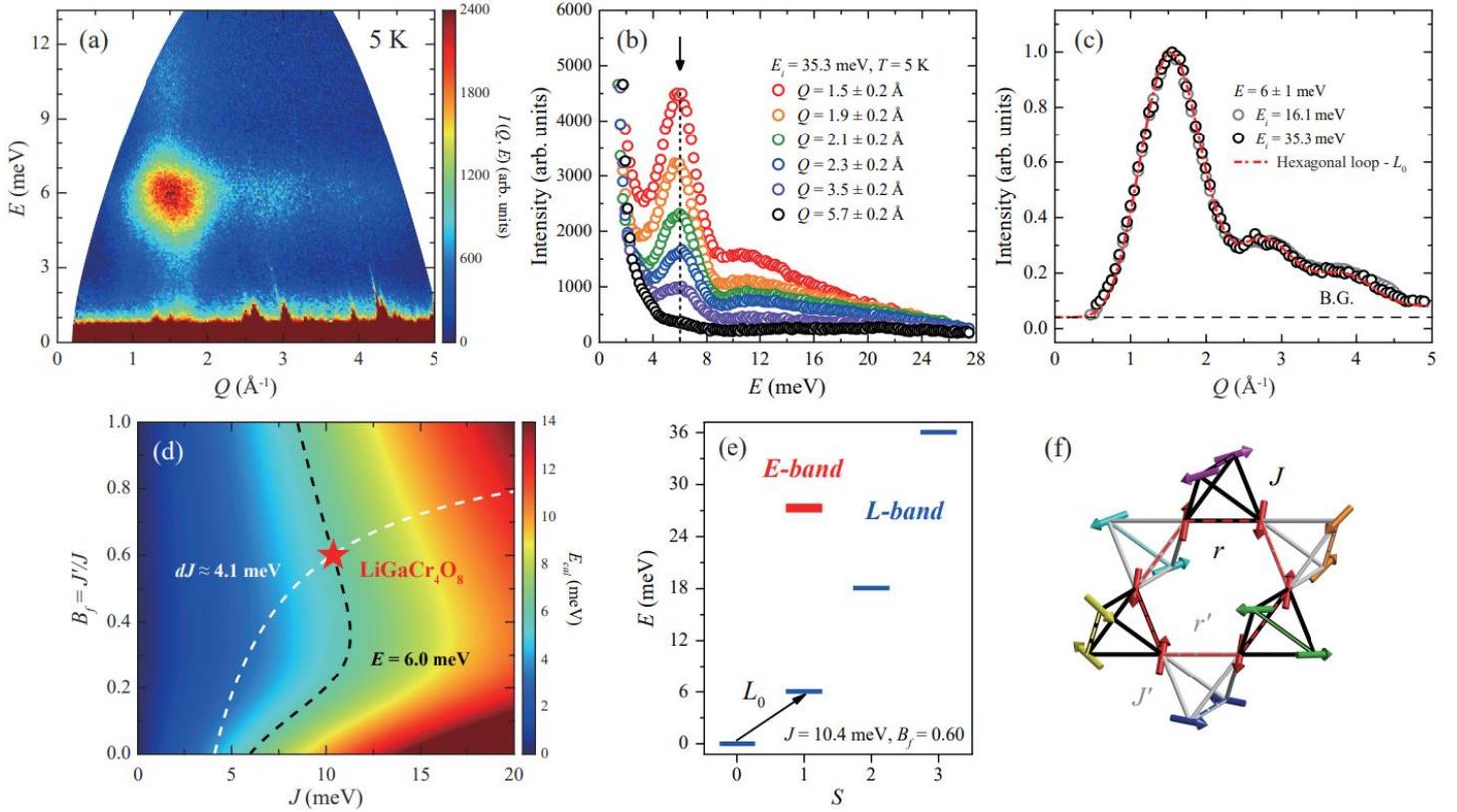

Fig. 2 (Color online) (a) Magnetic excitation spectra of LiGaCr$_4$O$_8$ at 5 K with incident neutron energy $E_i$ = 16.1 meV. (b) Energy cuts at indicated $Q$. The arrow indicates the resonance mode. (c) $Q$ dependence of the spin resonance measured with different incident energies (grey and black circles for $E_i$ = 16.1 and 35.3 meV respectively) and the calculated spin excitation $L_0$ for hexagonal spin loops (the red curve) with a constant background. Error bars are smaller than the symbols. (d) Calculated resonance energies versus magnetic interactions $J$ and breathing factors $B_f$, which are based on the spin Hamiltonian (1). The black and white curves are described in the text and their intersection gives $J$ = 10.4 meV and $B_f$ = 0.60 for LiGaCr$_4$O$_8$. (e) The calculated energy spectrum versus the total spin $S$ of the hexagonal loop using the above parameter set. The blue and red symbols indicate $L$- and $E$- band spin levels. (f) The sketch of a hexagonal spin loop on the lattice of LiGaCr$_4$O$_8$. The arrows marked by different colors stand for the spins belonging to different loops.

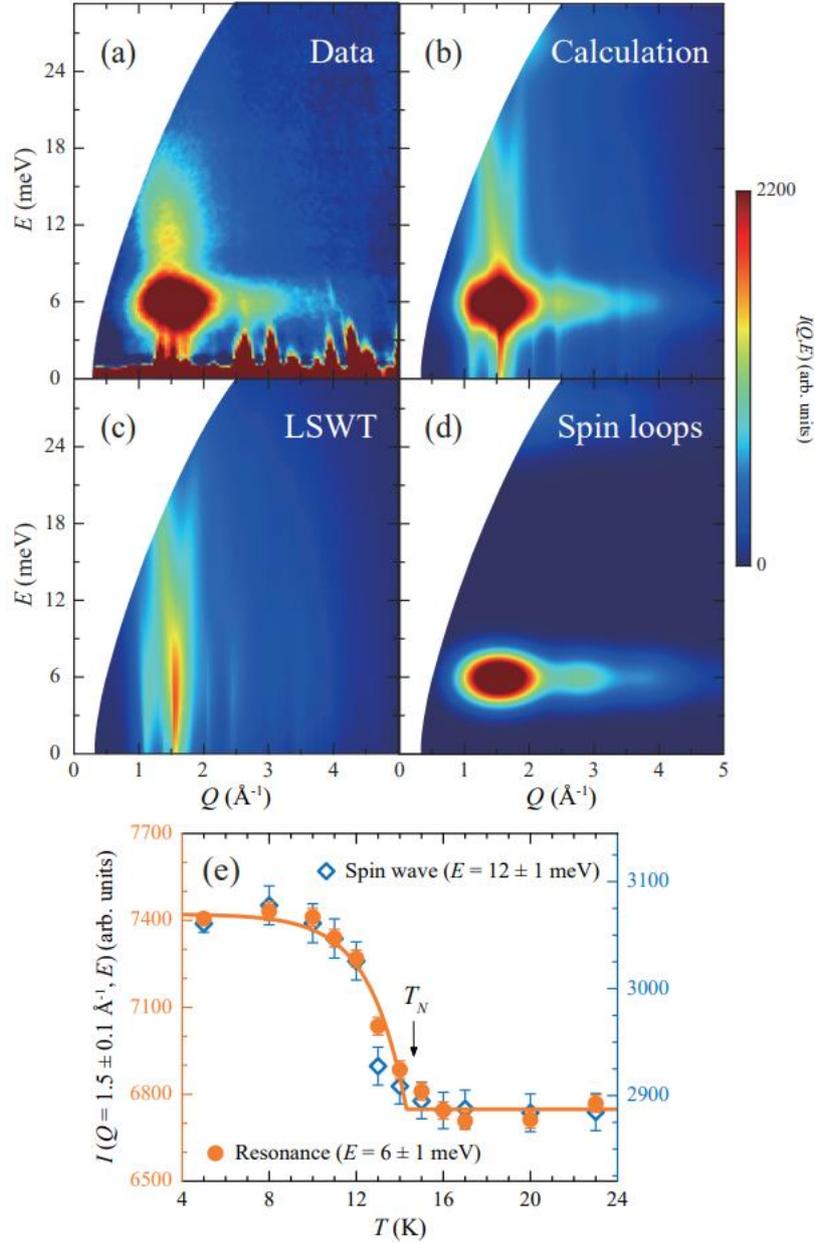

Fig. 3 (Color online) (a) Magnetic excitation spectra of LiGaCr$_4$O$_8$ at 5 K with incident neutron energy $E_i$ = 35.5 meV after subtracting the incoherent background at $Q$ = 6.4 Å$^{-1}$. (b) The total simulated scattering spectra combining the contributions from: (c) calculated spin wave spectra including two magnetic orders determined in Ref. [15] and magnetic form factor of Cr$^{3+}$ and (d) calculated hexagonal spin loops' excitations using the dynamical structure factor in Eq. (2). The same set of exchange parameters with $J$ = 10.4 meV and $J'$ = 6.2 meV are used in above calculations. (e) Temperature dependences of the spin resonance and spin wave.

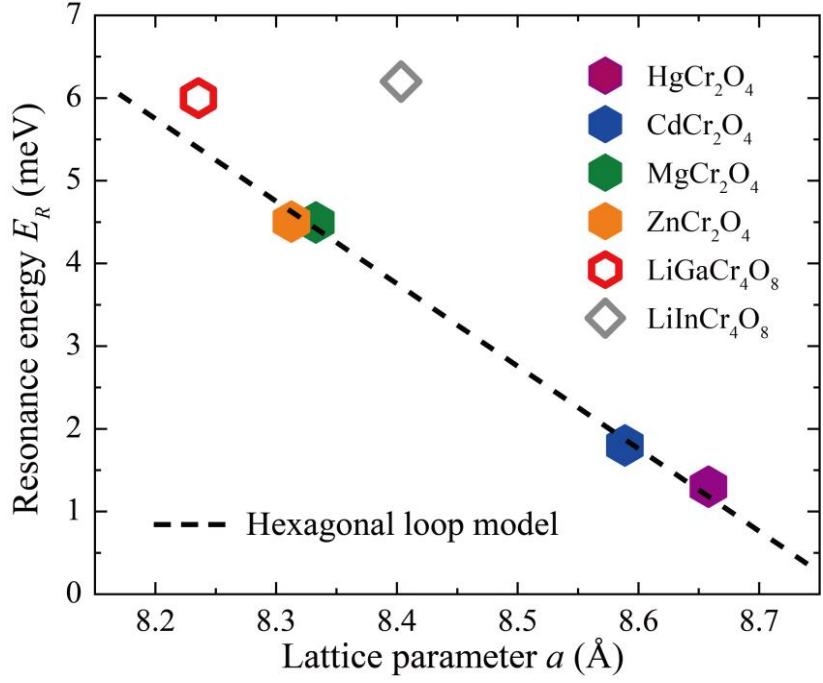

Fig. 4 (Color online) The relationship between resonance energies $E_R$ and lattice parameters $a$ of Cr-based spinel oxides. Energies of resonance are obtained from inelastic neutron results [6-10,34] and corresponding lattice parameters are determined via diffraction measurements [6,9,15,32-34]. The dashed line represents the calculated linear relation between $E_R$ and $a$ with $dE_R/da$ = -10 meV/Å using our quantum hexagonal loop model with the fixed breathing factor $B_f$ = 1 and the universal ratio of $dJ/dr \approx$ -40 meV/Å in chromium oxides [6,9,28].